\documentclass[prl,twocolumn,superscriptaddress,nofootinbib]{revtex4-1}
\usepackage{graphicx}
\usepackage{amsmath}
\usepackage{dcolumn}
\usepackage{color}
\usepackage{epstopdf}
 
\usepackage[caption=false]{subfig}

\begin{document} 

\title{Helicity-Flux-Driven Alpha Effect in Laboratory and Astrophysical Plasmas}
 
\author{F. Ebrahimi}
\affiliation{Center for Magnetic Self-Organization, Max Planck/Princeton Center for Plasma Physics, and Princeton Center for Heliospheric Physics,
Department of Astrophysical Sciences,
Princeton University, Princeton, NJ 08544
}   
\author{A. Bhattacharjee}
\affiliation{Center for Magnetic Self-Organization, Max Planck/Princeton Center for Plasma Physics, and Princeton Center for Heliospheric Physics,
Department of Astrophysical Sciences,
Princeton University, Princeton, NJ 08544
}
\affiliation{Princeton Plasma Physics Laboratory, Princeton, NJ 08543
}

\date{\today}
  
\begin{abstract}
The constraint imposed by magnetic helicity conservation on the alpha effect
is considered for both magnetically and flow dominated self-organizing plasmas. Direct
numerical simulations are presented for a dominant contribution to the alpha effect,
which can be cast in the functional form of a total divergence of an averaged helicity
flux, called the helicity-flux-driven alpha (H$\alpha$) effect. Direct numerical simulations of the H$\alpha$ effect are presented for two examples---the magnetically dominated toroidal plasma unstable to tearing modes, and the flow-dominated accretion disk.

\end{abstract}
\maketitle

Large-scale magnetic fields have been observed in widely different types of astrophysical objects, such as 
planets and stars, as well as accretion disks and galaxies.
The source of this magnetic field is the well-known dynamo effect,
which has stimulated an extensive search for models in which large-scale
magnetic fields are self-generated from turbulence and sustained despite the presence of
dissipation. The point of departure of most theoretical and computational studies of this
problem is magnetohydrodynamics (MHD), represented by intrinsically nonlinear
equations that describe the self-consistent evolution of a magnetized fluid. A standard
approach to the problem is mean-field theory, in which a fluctuation-induced electromotive
force (emf) parallel to the mean magnetic field is obtained from the vector
product of flow and magnetic field fluctuations. This is known as the alpha effect, which
holds a key to how large-scale magnetic fields may grow out of turbulence, beginning
from a seed field. 

 While kinematic dynamo
theory~\cite{moffat} predicts the existence of the alpha effect in astrophysical settings 
given a complex velocity field, its magnitude and saturation in a fully nonlinear,
self-consistent theory has been a subject of significant controversy. Since this problem
appears to be just as difficult as the problem of MHD turbulence, one constructive
approach is to examine the constraints imposed by rigorous conservation
laws of MHD on the functional form of the alpha effect. In this context, the role of
magnetic helicity, which is a ``rugged invariant'' of MHD turbulence~\cite{kraichnan67}, has received 
significant attention.~\cite{blackman2010}    

In this Letter, we revisit the role of magnetic helicity flux on the alpha effect.
We do so by considering two completely different physical examples from a common
perspective. The first example is a magnetically dominated self-organized toroidal plasma such as the reversed field pinch (RFP) in which the alpha effect is instrumental in converting one type of magnetic flux into another by the intervention of tearing instabilities while the total magnetic energy decays. The second
example is a flow-driven accretion disk, which too exhibits self-organization, and where
there is compelling evidence from several MHD simulations that a large-scale magnetic
field is produced and sustained by the nonlinear evolution of the magnetorotational
instability (MRI).~\cite{balbus91}  
The fluctuation-induced alpha effect due to the MRI results in the generation of large-scale magnetic field, which can cause MRI saturation~\cite{ebrahimi2009}.
We demonstrate by analysis and direct numerical simulations (DNS)
that \textit{in both cases a dominant contribution to the alpha effect} can be cast in the
functional form of a total divergence of an averaged helicity flux, which we call the
helicity-flux-driven alpha effect (hereafter simply referred to as the H$\alpha$ effect), and
is represented by the last term in Eq.~\ref{eq:yuan1} below. In the case of the RFP, the H$\alpha$ effect
reduces to ``hyperresistivity''~\cite{hyperresistivityBH86,boozer86,strauss86,finn2005}, often invoked by the MHD turbulence community
but demonstrated here to emerge from DNS of tearing modes. In the case of
the accretion disk, the H$\alpha$ effect leads to a new type of flux that is related to, but is more
complete than, the so-called Vishniac-Cho flux~\cite{vishniac_cho}, which was obtained by a heuristic
reduction of the nonlinear MHD equations, and has been invoked in recent astrophysical
dynamo studies.~\cite{sur2007,subramanian2004nonlinear}  
Here, we demonstrate from a global MRI
simulation~\cite{ebrahimi2009} that the H$\alpha$ effect plays a critical role in the self-generation of the large-scale
magnetic field. Viewed together, these two different physical applications reinforce the 
importance of the constraint Eq.~\ref{eq:yuan1}, derived rigorously from the MHD equations,
and the H$\alpha$ effect that emerges from it.

We begin with a discussion of the constraint equation, which was first discussed
in ~\cite{hyperresistivityBH86} in the context of the RFP, and later in a form more relevant for astrophysical
applications in~\cite{bh_yuan95} as well as~\cite{blackmanfield}. Using the equations for the time-derivative of the
vector potential A and magnetic field B, given by Maxwell’s equations 
$\partial \textbf{A}/\partial t = -\textbf E - \nabla \phi$ and 
$\partial \textbf{B}/\partial t = -\nabla \times \textbf E$, where $\phi$
 is the electrostatic potential, we obtain
\begin{equation}
 \frac{1}{2} \frac{\partial (\textbf A \cdot \textbf B)}{\partial t} + \frac{1}{2} \nabla \cdot \Gamma_k =  - \textbf E \cdot \textbf B. 
\label{eq:hel}
\end{equation} 
where $\Gamma_k = (-\textbf A \times \textbf E  + \textbf A \times \nabla \phi) = -2 \textbf A \times \textbf E  - \textbf A \times \partial\textbf A/\partial t$ is defined as 
the total magnetic helicity flux. All variables are decomposed as $f=<f>+\widetilde f$, where $<f> = \overline{f}$ is the \textit {mean} component (where $< \cdots >$ denotes the azimuthal and axial average), and $\widetilde f$ is the fluctuating component. It can be shown that,~\cite{bh_yuan95,blackmanfield}
\begin{equation}
\frac{1}{2}\frac{\partial <\widetilde{\textbf{A}} \cdot \widetilde{\textbf{B}}  >}{\partial t} + \frac{1}{2}\nabla \cdot <\Gamma_k> = -  <\widetilde \textbf E \cdot \widetilde \textbf B>. 
\label{eq:helfluc}
\end{equation}
We now use the perturbed form of Ohm’s law for a resistive MHD plasma, $\widetilde{\textbf E} = -\overline{\textbf V} \times \widetilde \textbf B - \widetilde \textbf V \times \overline{\textbf B} + \eta \widetilde{\textbf J}$, which implies that
$\varepsilon_{emf} \cdot \overline{\textbf B}=  < \widetilde{\textbf{V}} \times  \widetilde{\textbf{B}}  > \cdot \overline{\textbf B}
 = - \eta <\widetilde \textbf J \cdot \widetilde \textbf B> + <\widetilde{\textbf E} \cdot \widetilde \textbf B>$, which when combined
with Eq.~\ref{eq:helfluc}, yields the exact result
\begin{equation}
\varepsilon_{emf} \cdot \overline{\textbf B} = -\eta <\widetilde{\textbf{J}} \cdot \widetilde{\textbf{B}} > -\frac{1}{2} \frac{\partial}{\partial t}<\widetilde{\textbf{A}} \cdot \widetilde{\textbf{B}} > 
+ H{\alpha}
\label{eq:yuan1} 
\end{equation}
where 
\begin{equation}
H{\alpha} =  -\frac{1}{2}\nabla \cdot <\Gamma_k> = \nabla \cdot [<\widetilde{\textbf{A}} \times \widetilde{\textbf{E}} >+ \frac{1}{2} <\widetilde{\textbf{A}} \times \frac{\partial}{\partial t}\widetilde{\textbf{A}} >]. 
\label{eq:yuan2} 
\end{equation}
Equations~\ref{eq:yuan1} and~\ref{eq:yuan2} serve as the point of departure for both of our examples, discussed below. Specifically, for both examples, the H$\alpha$ effect will be calculated using Eq.~\ref{eq:yuan2}.
We employ the nonlinear, resistive MHD code, DEBS, which solves the single fluid MHD equations in doubly 
periodic $(r,\phi,z)$ cylindrical geometry \cite{Debs,ebrahimi2009}.
We use the same normalization as in~\cite{Debs,ebrahimi2009}, where time, radius and velocity are normalized to the outer radius \textit{a}, the resistive diffusion time $\tau_R = a^2/\mu_0\eta$, and the Alfv\'en velocity $V_A = B_0/\sqrt(\mu_0 \rho_0)$, respectively $(B_0$ and $\rho_0$ are the values on axis).
The dimensionless parameters, $S= \tau_R V_A/a$ and $P_m$, 
are the Lundquist
number and  the magnetic Prandtl number (the ratio of viscosity to resistivity), respectively. 
For magnetically dominated simulations a force-free initial condition is used~\cite{ebrahimithesis}. For the flow dominated MRI simulations, the initial state satisfies the equilibrium force balance condition $ \frac{\beta_0}{2}\nabla p  = \rho V_{\phi}^2/r$, where $\beta_0 = 2 \mu_0 P_0/B_0^2$ is the beta normalized 
to the axis value, and 
the initial pressure and density profiles are assumed to be radially uniform and nonstratified. Pressure and
density are evolved; however, they remain fairly uniform during the computations. In these computations, a mean Keplerian profile ($V_{\phi} \propto r^{-1/2}$) is maintained in time by an external ad-hoc force in the momentum equation.
 The boundary conditions in 
the radial direction are as are appropriate for dissipative 
MHD with a perfectly conducting boundary: the tangential electric 
field, the normal component of the magnetic field, and the normal 
component of the velocity vanish, and the tangential component of 
the velocity is the rotational velocity of the wall.  
The azimuthal $(\phi)$ and axial $(z)$ directions are periodic. The boundary conditions on the magnetic field (and the vector potential) keep the volume-integrated magnetic helicity gauge-invariant.~\footnote{Under a gauge transformation $\textbf A \rightarrow \textbf A+\nabla f$, the volume-integrated helicity becomes $ \int \textbf A \cdot \textbf B dV + \int f \textbf B \cdot d\textbf{s}$, and magnetic helicity is therefore gauge-invariant when $\textbf B \cdot \hat n= 0$ (as in our simulations). Also, in our tearing simulations with an initial force-free equilibrium, no loop voltages are applied to sustain the current, so the total helicity, as defined, remains gauge-invariant.}
       
In magnetically dominated laboratory configurations like the RFP~\cite{ji95} and spheromaks~\cite{jarboe1994review,spheromak}, the importance
of magnetic helicity is well recognized due to the efficiency of the Taylor relaxation
process~\cite{taylor74}. In these configurations, it has been experimentally demonstrated that a
turbulent plasma relaxes to a state of minimum energy subject to the conservation of
total magnetic helicity~\citep{ji95,spheromak}. There is strong evidence from experiments as well as 
simulations~\cite{ebrahimithesis} that tearing instabilities resonant with rational surfaces within the plasma
play an important role in the relaxation process. The tearing fluctuations contribute to the 
emf $\varepsilon_{emf}$ , which converts poloidal flux 
to toroidal flux. 

To calculate $\varepsilon_{emf}$ due to tearing instabilities, we consider a general cylindrical
equilibrium magnetic field  $\overline{\textbf B} = B_z(r) \hat{z} + B_{\phi}(r) \hat{\phi}$ which is 
subject to perturbations of the form 
$\tilde{f} (r, \phi,z,t) = \tilde{f}(r)$ 
exp$(\gamma t + i m \phi - i n z/R)$ in cylinder of radius a and
periodicity length 2 $\pi$ R along z.
In this geometry, reconnection driven by tearing instabilities tend to occur at mode-rational surfaces located at $r=r_s$, where $q = rB_z(r)/RB_{\phi}(r)= m/n$ and m and n are positive integers. At these resonant surfaces, the parallel wave number 
vector vanishes, that is $\textbf{k} \cdot \textbf{B} = m B_{\phi}/r - n B_z/R = 0$. 

For reconnecting instabilities, we adopt the standard tearing ordering~\cite{CGJ}
$\gamma \propto \eta^{3/5}$, $\gamma$ $\rightarrow$ $\epsilon^3$,  
$\eta$ $\rightarrow$ $\epsilon^5$,  $(r-r_s)$ $\rightarrow$ $\epsilon^2 x$, where
 $\epsilon$ is a small parameter. Using this ordering, it can be
shown that to leading order Eq.~\ref{eq:yuan1} reduces to $-\nabla \cdot <(\widetilde \textbf A \cdot \overline{\textbf B}) \widetilde \textbf V> $, which can be written in terms of the perturbed radial displacement ($\Xi$) and the lowest-order perturbed radial magnetic field ($\Psi_0$)~\cite{hyperresistivityBH86,hameiri_bh87}. One obtains 
\begin{equation}
\varepsilon_{emf} \cdot \overline{\textbf B} = \nabla \cdot (\kappa^2 \nabla \frac{\overline{\textbf J} \cdot \overline{\textbf B}}{\overline{B}^2})
\label{eq:hyper}
\end{equation}
where $\kappa^2$ is a positive-definite function. Equation~\ref{eq:hyper}, well established in the literature~\cite{hyperresistivityBH86,boozer86,strauss86,finn2005,hameiri_bh87}, is the magnetic helicity conserving form of the alpha effect, which through hyperresistivity relates the turbulent emf to the gradient of mean parallel current density (the free energy for tearing instability).  Here, we include full radial structure of $\Xi$ and further extend the derivation of $\kappa^2$ to obtain the radial extent and the structure of hyperresistivity around the reconnection surface in terms of Hermite polynomials to obtain,
\begin{equation}
\kappa^2 = \frac{B^2}{k r_s} \eta Q \Psi^2_0 \zeta(r), 
\label{eq:k2}
\end{equation}
\begin{equation}
\begin{split}
\zeta(r) &= \mathrm{exp}[-(Q^{-1/2}(r-r_s))^2/2] \Sigma_m a_{2m} H_{2m}[Q^{-1/4}(r-r_s)],\\
a_{2m} &= \frac{ 2^{1/2} }{4^m m!}
[\frac{\frac{1}{Q^{3/2}}(4m+1)}{(4m+1)(4m+1) + S_q}],\\
\end{split}
\label{eq:a2m}
\end{equation}
where Q and $S_q $ are the normalized growth rate and shear factor, defined as  $Q = \gamma/(l^4 Q_R)$ and $S_q =(\frac{2 a}{R q^{\prime}})^2 $, where prime indicates radial derivative ($Q_R = (\frac{m^2 \eta F^{\prime 2} B^2}{\rho r_s^2})^{1/3}$, l$=(L_R/r_s)^{1/5}$, $L_R = (\eta/Q_R)^{1/2}$, $F^{\prime} = - \frac{q^{\prime}B_{\theta}}{q B}$). We note that
only the even parity solution for the perturbed radial displacement contributes to 
H$\alpha$ under the “constant-psi” approximation, represented by the constant $\Psi_0$. (We adopt here
the notation of~\cite{ebrahimi2008}.)

\begin{figure}  
 \includegraphics[]{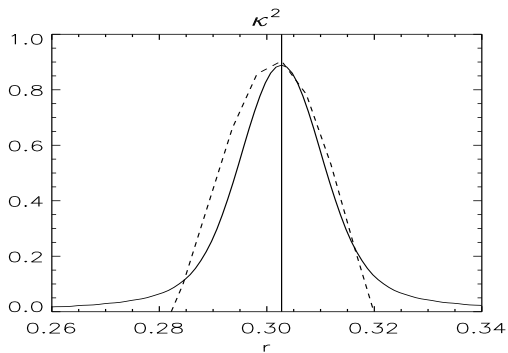}   
\caption{The radial structure of hyperresistivity, $\kappa^2$ for a single tearing mode with S =$10^6$. The solid line denotes the analytical result, while the dashed line shows the numerical result in the vicinity of the reconnecting surface where linear theory is applicable. The reconnecting surface is shown at $r=r_s$ with the vertical line.} 
\label{fig:fig1}
\end{figure}  

For a reconnecting tearing mode (m=1, k=1.8) resonant at r=0.303, with a mean parallel current, $\overline{\textbf J} \cdot \overline{\textbf B} /\overline{B}^2 =3.2 (1-r^3)$, the hyperresistivity, $\kappa^2$, 
in the inner reconnection layer 
can directly be obtained analytically from Eqs.~\ref{eq:k2} and \ref{eq:a2m}, which shows a radial extent of about 0.05 (here we have
normalized r to the minor radius a).
In Fig.~\ref{fig:fig1}, we compare the analytical result
with the numerical result obtained from a single tearing mode computation.  
Both results demonstrate the positivity of $\kappa^2$, required by conservation laws.~\cite{hyperresistivityBH86,boozer86} The dashed line represents the numerical result from the leading term in Eq.~\ref{eq:yuan1}, and the solid line the analytical result. We should 
note that 
the computations are in the viscous-resistive regime with $S= 10^6$ and Pm=1, which results in a larger 
radial extent of $\kappa^2$  than the analytical result. We have also computed $\kappa^2$ from multiple
nonlinear tearing modes, which shows a broadening from multiple modes. Thus, our DNS confirm for the first time the explicit functional form of Eq.~\ref{eq:hyper}. The predictions of Eq.~\ref{eq:hyper} for the RFP are well-known----the mean-field saturated
states in these magnetically dominated self-organized plasmas are Taylor-like ~\cite{hyperresistivityBH86,boozer86,strauss86},
consistent with experimental observations~\citep{ji95,spheromak}. Taylor relaxation theory has also been adapted to explain the heating of the solar corona~\cite{priest84}. The possible role of hyperresistivity in this process has also been examined.~\cite{vanballegoo2008}     
\begin{figure} 
\includegraphics[]{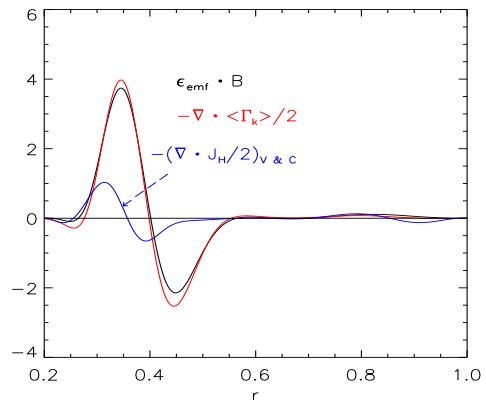}
\caption{Dynamo term $\varepsilon_{emf} \cdot \overline{\textbf B}$, total divergence form of fluctuation-induced 
helicity flux given in equation (4) $- \nabla \cdot <\Gamma_k>/2 $, and the divergence form of helicity flux given by Vishniac \& Cho, during m=1 MRI mode nonlinear saturation.}
\vspace{-4mm}
\label{fig:fig2} 
\end{figure}       

We now consider the H$\alpha$ effect in the context of the accretion disk, which is flow-driven
and dominated by the MRI. In contrast to the magnetically dominated RFP,
where the primary source of free energy is the parallel current density, the primary source
of free energy is the accretion disk is the flow shear. Motivated in part by 
the considerations of~\cite{bh_yuan95}, Vishniac and Cho~\cite{vishniac_cho} proposed a form of helicity-flux-driven flux of the form $(- \nabla \cdot J_H/2)_{VC}$ where $J_H =   <(\widetilde{\textbf{E}} - \nabla \widetilde \phi) \times \widetilde{\textbf{A}}> = - 2<(\widetilde{\textbf{E}} \times \nabla \widetilde \phi)> \tau_c$  where $\tau_c$ is the eddy correlation time. We demonstrate below by DNS that $(- \nabla \cdot J_H/2)_{VC}$ 
underestimates significantly the role of the H$\alpha$ effect. As is often standard practice in
zero-net-flux MRI simulations, we begin with an initial magnetic field of the form $B_z=sin(2 \pi (r-r1)/(r2-r1))/r$ and $B_{\phi} =0$, driven by an azimuthal mean Keplerian flow ($V_{\phi} = V_0 r^{-1/2}$), where $r1$, $r2$ and $V_0$ are the inner and outer radii and the magnitude of mean flow on axis, respectively. Figure~\ref{fig:fig2} shows the result of a single non-axisymmetric m=1 mode
computation of the MRI in which the various terms in Eq.~\ref{eq:yuan1} are calculated numerically
for magnetic Reynolds number Rm = $S V_0/V_A$ =1600 (with Pm =1), in a nonlinearly saturated state. It is seen that the
term on the left-hand-side, given by $\varepsilon_{emf} \cdot \overline{\textbf B}$ is balanced almost entirely by H$\alpha$, and the
contribution of the other two terms are small. For comparison, we also compute the
Vishniac-Cho flux, which is much smaller than H$\alpha$. This is because the perturbed electric field computed in ~\cite{vishniac_cho} is approximated by the relation $\widetilde{\textbf{E}} = - \widetilde \textbf V \times \overline{\textbf B}$, which is incomplete. Note that our computation of the H$\alpha$ effect begins with the exact Eq.~\ref{eq:yuan2}, which differs from the approximate equations used in other studies~\cite{subramanian2004nonlinear,sordo2013,shapovalov2013} which have also identified additional contributions to the Vishniac-Cho flux.  
\begin{figure}
 \includegraphics[]{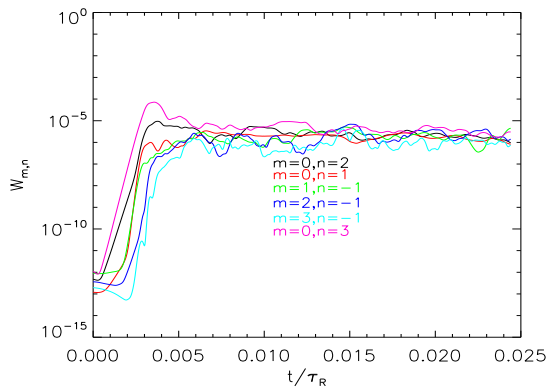} 
\caption{Magnetic energy $W_{m,n}=1/2 \int{\widetilde B_{r(m,n})^2 dr^3}$ vs time for
different MRI modes for a fully nonlinear 
computations.  }
\vspace{-4mm}
\label{fig:fig3}
\end{figure} 

\begin{figure}
\includegraphics[]{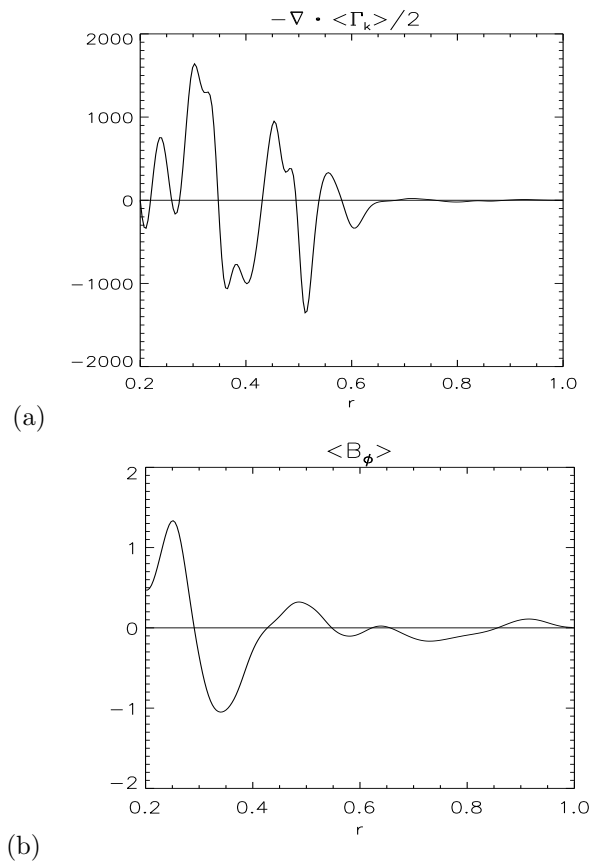}
\caption{The radial profiles of (a) H$\alpha$ for all the MRI modes (b) large-scale mean azimuthal magnetic field generated, averaged over three points in time (t/$\tau_R$=0.006,0.0062, 0.0065).}
\vspace{-4mm}
\label{fig:fig4}
\end{figure}

We have also carried out DNS with multiple modes, leading to MRI turbulence.
 Fully nonlinear computations start with a Keplerian flow 
profile and zero net-flux with $P_m=2$, $Rm=3400$, $\beta_0=10^5$, 
 and radial, azimuthal and axial 
resolutions  $n_r$=220, $0<m<43$ and $-43<n<43$, respectively. 
The radial magnetic energy for MRI modes is shown in 
Fig.~\ref{fig:fig3}. As shown, the axisymmetric m=0 modes as well as the
non-axisymmetric modes (m=1,2,3 are shown) grow robustly 
in the linear regime and
saturate non-linearly. The nonlinear saturated state 
exhibits significant time-dependence,
including the second term on the right-hand-side of Eq.~\ref{eq:yuan1}. In Fig.~\ref{fig:fig4}, we show the
time-averaged H$\alpha$  profile over three instants of time, 
and the averaged (along $\phi$ and z directions) $B_{\phi}$ profile.
Since the initial state had no $B_{\phi}$, it is clear that the H$\alpha$
 effect plays a dominant role
in producing a large-scale azimuthal field which, in turn, contributes to the nonlinear
saturation of the MRI~\cite{ebrahimi2009}. (The Vishniac-Cho flux continues to be much smaller than
H$\alpha$ under these circumstances.) 

In conclusion, we have demonstrated by DNS in global geometry that the  
H$\alpha$
effect, which is a rigorous consequence of magnetic helicity conservation in a turbulent
plasma, plays a dominant role in magnetically as well as flow-driven self-organization.
While the case of the magnetically dominated RFP, which has been the subject of numerous direct experimental observations, is relatively well understood, there remain many open questions in the case of the flow-driven MRI dynamo. Viewing both of these problems from a common perspective enables us to emphasize the importance of tracking the flow of magnetic helicity in quantitative measurements of the dynamo effect, which we have performed in this paper using DNS beginning with the exact equations (3) and (4). It is also clear that the averaging process as well as the boundary conditions used can play a subtle role in such quantitative measurements. Such considerations might play an important role in resolving important differences in results between shearing box~\citep{lesurdynamo,stratified2010,simon2011} and global MRI simulations.
    
\acknowledgements 
This work was supported by the National Science Foundation grant 
No. 0962244 and by DOE, DE-FG02-12ER55142, DE-FG02-07ER46372, and CMSO. This work was also 
facilitated by the Max-Planck/Princeton Center for Plasma Physics. We thank J. Stone for several stimulating discussions and F. E. appreciates valuable discussions on the numerical techniques used in this paper with Dalton Schnack.


\begin{thebibliography}{27}
\expandafter\ifx\csname natexlab\endcsname\relax\def\natexlab#1{#1}\fi
\expandafter\ifx\csname bibnamefont\endcsname\relax
  \def\bibnamefont#1{#1}\fi
\expandafter\ifx\csname bibfnamefont\endcsname\relax
  \def\bibfnamefont#1{#1}\fi
\expandafter\ifx\csname citenamefont\endcsname\relax
  \def\citenamefont#1{#1}\fi
\expandafter\ifx\csname url\endcsname\relax
  \def\url#1{\texttt{#1}}\fi
\expandafter\ifx\csname urlprefix\endcsname\relax\def\urlprefix{URL }\fi
\providecommand{\bibinfo}[2]{#2}
\providecommand{\eprint}[2][]{\url{#2}}

\bibitem[{\citenamefont{{Moffatt}}(1978)}]{moffat}
\bibinfo{author}{\bibfnamefont{H.~K.} \bibnamefont{{Moffatt}}},
  \emph{\bibinfo{title}{{Magnetic field generation in electrically conducting
  fluids}}} (\bibinfo{year}{1978}).

\bibitem[{\citenamefont{{Kraichnan}}(1967)}]{kraichnan67}
\bibinfo{author}{\bibfnamefont{R.~H.} \bibnamefont{{Kraichnan}}},
  \bibinfo{journal}{Physics of Fluids} \textbf{\bibinfo{volume}{10}},
  \bibinfo{pages}{1417} (\bibinfo{year}{1967}).

\bibitem[{\citenamefont{{Blackman}}(2010)}]{blackman2010}
\bibinfo{author}{\bibfnamefont{See, for instance, the recent overview by E.~G.} \bibnamefont{{Blackman}}},
  \bibinfo{journal}{Astronomische Nachrichten} \textbf{\bibinfo{volume}{331}},
  \bibinfo{pages}{101} (\bibinfo{year}{2010}), \eprint{0911.2315}.

\bibitem[{\citenamefont{Balbus and Hawley}(1991)}]{balbus91}
\bibinfo{author}{\bibfnamefont{S.~A.} \bibnamefont{Balbus}} \bibnamefont{and}
  \bibinfo{author}{\bibfnamefont{J.~F.} \bibnamefont{Hawley}},
  \bibinfo{journal}{ApJ} \textbf{\bibinfo{volume}{376}},
  \bibinfo{pages}{214} (\bibinfo{year}{1991}).

\bibitem[{\citenamefont{Ebrahimi et~al.}(2009)\citenamefont{Ebrahimi, Prager,
  and Schnack}}]{ebrahimi2009}
\bibinfo{author}{\bibfnamefont{F.}~\bibnamefont{Ebrahimi}},
  \bibinfo{author}{\bibfnamefont{S.~C.} \bibnamefont{Prager}},
  \bibnamefont{and} \bibinfo{author}{\bibfnamefont{D.~D.}
  \bibnamefont{Schnack}}, \bibinfo{journal}{Astrophys. J.}
  \textbf{\bibinfo{volume}{698}}, \bibinfo{pages}{233} (\bibinfo{year}{2009}).
 
\bibitem[{\citenamefont{{Bhattacharjee} and
  {Hameiri}}(1986)}]{hyperresistivityBH86}
\bibinfo{author}{\bibfnamefont{A.}~\bibnamefont{{Bhattacharjee}}}
  \bibnamefont{and}
  \bibinfo{author}{\bibfnamefont{E.}~\bibnamefont{{Hameiri}}},
  \bibinfo{journal}{Phys. Rev. Lett.} \textbf{\bibinfo{volume}{57}},
  \bibinfo{pages}{206} (\bibinfo{year}{1986}).
 
\bibitem[{\citenamefont{Boozer}(1986)}]{boozer86}
\bibinfo{author}{\bibfnamefont{A.~H.} \bibnamefont{Boozer}},
  \bibinfo{journal}{J. Plasma Phys.} \textbf{\bibinfo{volume}{35}},
  \bibinfo{pages}{133} (\bibinfo{year}{1986}).

\bibitem[{\citenamefont{{Strauss}}(1986)}]{strauss86}
\bibinfo{author}{\bibfnamefont{H.~R.} \bibnamefont{{Strauss}}},
  \bibinfo{journal}{Physics of Fluids} \textbf{\bibinfo{volume}{29}},
  \bibinfo{pages}{3668} (\bibinfo{year}{1986}).

\bibitem[{\citenamefont{{Finn}}(2005)}]{finn2005}
\bibinfo{author}{\bibfnamefont{J.~M.} \bibnamefont{{Finn}}},
  \bibinfo{journal}{Physics of Plasmas} \textbf{\bibinfo{volume}{12}},
  \bibinfo{pages}{092313} (\bibinfo{year}{2005}).

\bibitem[{\citenamefont{{Vishniac} and {Cho}}(2001)}]{vishniac_cho}
\bibinfo{author}{\bibfnamefont{E.~T.} \bibnamefont{{Vishniac}}}
  \bibnamefont{and} \bibinfo{author}{\bibfnamefont{J.}~\bibnamefont{{Cho}}},
  \bibinfo{journal}{ApJ} \textbf{\bibinfo{volume}{550}}, \bibinfo{pages}{752}
  (\bibinfo{year}{2001}).

\bibitem[{\citenamefont{{Sur} et~al.}(2007)\citenamefont{{Sur}, {Shukurov}, and
  {Subramanian}}}]{sur2007}
\bibinfo{author}{\bibfnamefont{S.}~\bibnamefont{{Sur}}},
  \bibinfo{author}{\bibfnamefont{A.}~\bibnamefont{{Shukurov}}},
  \bibnamefont{and}
  \bibinfo{author}{\bibfnamefont{K.}~\bibnamefont{{Subramanian}}},
  \bibinfo{journal}{MNRAS} \textbf{\bibinfo{volume}{377}}, \bibinfo{pages}{874}
  (\bibinfo{year}{2007}).

\bibitem[{\citenamefont{Subramanian and
  Brandenburg}(2004)}]{subramanian2004nonlinear}
\bibinfo{author}{\bibfnamefont{K.}~\bibnamefont{Subramanian}} \bibnamefont{and}
  \bibinfo{author}{\bibfnamefont{A.}~\bibnamefont{Brandenburg}},
  \bibinfo{journal}{Phys. Rev. Lett.} \textbf{\bibinfo{volume}{93}},
  \bibinfo{pages}{205001} (\bibinfo{year}{2004}).

\bibitem[{\citenamefont{{Bhattacharjee} and {Yuan}}(1995)}]{bh_yuan95}
\bibinfo{author}{\bibfnamefont{A.}~\bibnamefont{{Bhattacharjee}}}
  \bibnamefont{and} \bibinfo{author}{\bibfnamefont{Y.}~\bibnamefont{{Yuan}}},
  \bibinfo{journal}{ApJ} \textbf{\bibinfo{volume}{449}}, \bibinfo{pages}{739}
  (\bibinfo{year}{1995}).

\bibitem[{\citenamefont{{Blackman} and {Field}}(2000)}]{blackmanfield}
\bibinfo{author}{\bibfnamefont{E.~G.} \bibnamefont{{Blackman}}}
  \bibnamefont{and} \bibinfo{author}{\bibfnamefont{G.~B.}
  \bibnamefont{{Field}}}, \bibinfo{journal}{ApJ}
  \textbf{\bibinfo{volume}{534}}, \bibinfo{pages}{984} (\bibinfo{year}{2000}).

\bibitem[{\citenamefont{Schnack et~al.}(1987)\citenamefont{Schnack, Barnes,
  Mikic, Harned, and Caramana}}]{Debs}
\bibinfo{author}{\bibfnamefont{D.~D.} \bibnamefont{Schnack}},
  \bibinfo{author}{\bibfnamefont{D.~C.} \bibnamefont{Barnes}},
  \bibinfo{author}{\bibfnamefont{Z.}~\bibnamefont{Mikic}},
  \bibinfo{author}{\bibfnamefont{D.~S.} \bibnamefont{Harned}},
  \bibnamefont{and} \bibinfo{author}{\bibfnamefont{E.~J.}
  \bibnamefont{Caramana}}, \bibinfo{journal}{J. Comput. Phys.}
  \textbf{\bibinfo{volume}{70}}, \bibinfo{pages}{330} (\bibinfo{year}{1987}).

\bibitem[{\citenamefont{Ebrahimi}(2003)}]{ebrahimithesis}
\bibinfo{author}{\bibfnamefont{F.}~\bibnamefont{Ebrahimi}},
  \emph{\bibinfo{title}{Ph.D thesis, Nonlinear magnetohydrodynamics of AC
  helicity injection}} (\bibinfo{year}{2003}).

\bibitem[{\citenamefont{{Ji} et~al.}(1995)\citenamefont{{Ji}, {Prager}, and
  {Sarff}}}]{ji95}
\bibinfo{author}{\bibfnamefont{H.}~\bibnamefont{{Ji}}},
  \bibinfo{author}{\bibfnamefont{S.~C.} \bibnamefont{{Prager}}},
  \bibnamefont{and} \bibinfo{author}{\bibfnamefont{J.~S.}
  \bibnamefont{{Sarff}}}, \bibinfo{journal}{Phys. Rev. Lett.}
  \textbf{\bibinfo{volume}{74}}, \bibinfo{pages}{2945} (\bibinfo{year}{1995}).

\bibitem[{\citenamefont{Jarboe}(1994)}]{jarboe1994review}
\bibinfo{author}{\bibfnamefont{T.~R.} \bibnamefont{Jarboe}},
  \bibinfo{journal}{Plasma Physics and Controlled Fusion}
  \textbf{\bibinfo{volume}{36}}, \bibinfo{pages}{945} (\bibinfo{year}{1994}).

\bibitem[{\citenamefont{{Cothran} et~al.}(2009)\citenamefont{{Cothran},
  {Brown}, {Gray}, {Schaffer}, and {Marklin}}}]{spheromak}
\bibinfo{author}{\bibfnamefont{C.~D.} \bibnamefont{{Cothran}}},
  \bibinfo{author}{\bibfnamefont{M.~R.} \bibnamefont{{Brown}}},
  \bibinfo{author}{\bibfnamefont{T.}~\bibnamefont{{Gray}}},
  \bibinfo{author}{\bibfnamefont{M.~J.} \bibnamefont{{Schaffer}}},
  \bibnamefont{and}
  \bibinfo{author}{\bibfnamefont{G.}~\bibnamefont{{Marklin}}},
  \bibinfo{journal}{Phys. Rev. Lett.} \textbf{\bibinfo{volume}{103}},
  \bibinfo{eid}{215002} (\bibinfo{year}{2009}).

\bibitem[{\citenamefont{{Taylor}}(1974)}]{taylor74}
\bibinfo{author}{\bibfnamefont{J.~B.} \bibnamefont{{Taylor}}},
  \bibinfo{journal}{Physical Review Letters} \textbf{\bibinfo{volume}{33}},
  \bibinfo{pages}{1139} (\bibinfo{year}{1974}).

\bibitem[{\citenamefont{{Coppi} et~al.}(1966)\citenamefont{{Coppi}, {Greene},
  and {Johnson}}}]{CGJ}
\bibinfo{author}{\bibfnamefont{B.}~\bibnamefont{{Coppi}}},
  \bibinfo{author}{\bibfnamefont{J.~M.} \bibnamefont{{Greene}}},
  \bibnamefont{and} \bibinfo{author}{\bibfnamefont{J.~L.}
  \bibnamefont{{Johnson}}}, \bibinfo{journal}{Nuclear Fusion}
  \textbf{\bibinfo{volume}{6}}, \bibinfo{pages}{101} (\bibinfo{year}{1966}).

\bibitem[{\citenamefont{{Hameiri} and {Bhattacharjee}}(1987)}]{hameiri_bh87}
\bibinfo{author}{\bibfnamefont{E.}~\bibnamefont{{Hameiri}}} \bibnamefont{and}
  \bibinfo{author}{\bibfnamefont{A.}~\bibnamefont{{Bhattacharjee}}},
  \bibinfo{journal}{Phys. Fluids} \textbf{\bibinfo{volume}{30}},
  \bibinfo{pages}{1743} (\bibinfo{year}{1987}).

\bibitem[{\citenamefont{Ebrahimi et~al.}(2008)\citenamefont{Ebrahimi, Mirnov,
  and Prager}}]{ebrahimi2008}
\bibinfo{author}{\bibfnamefont{F.}~\bibnamefont{Ebrahimi}},
  \bibinfo{author}{\bibfnamefont{V.~V.} \bibnamefont{Mirnov}},
  \bibnamefont{and} \bibinfo{author}{\bibfnamefont{S.~C.}
  \bibnamefont{Prager}}, \bibinfo{journal}{Phys. Plasmas}
  \textbf{\bibinfo{volume}{15}}, \bibinfo{pages}{055701}
  (\bibinfo{year}{2008}).

\bibitem[{\citenamefont{{Heyvaerts} and {Priest}}(1984)}]{priest84}
\bibinfo{author}{\bibfnamefont{J.}~\bibnamefont{{Heyvaerts}}} \bibnamefont{and}
  \bibinfo{author}{\bibfnamefont{E.~R.} \bibnamefont{{Priest}}},
  \bibinfo{journal}{Astronomy \& Astrophysics} \textbf{\bibinfo{volume}{137}},
  \bibinfo{pages}{63} (\bibinfo{year}{1984}).

\bibitem[{\citenamefont{{van Ballegooijen} and
  {Cranmer}}(2008)}]{vanballegoo2008}
\bibinfo{author}{\bibfnamefont{A.~A.} \bibnamefont{{van Ballegooijen}}}
  \bibnamefont{and} \bibinfo{author}{\bibfnamefont{S.~R.}
  \bibnamefont{{Cranmer}}}, \bibinfo{journal}{ApJ}
  \textbf{\bibinfo{volume}{682}}, \bibinfo{pages}{644} (\bibinfo{year}{2008}).

\bibitem[{\citenamefont{{Del Sordo} et~al.}(2013)\citenamefont{{Del Sordo},
  {Guerrero}, and {Brandenburg}}}]{sordo2013}
\bibinfo{author}{\bibfnamefont{F.}~\bibnamefont{{Del Sordo}}},
  \bibinfo{author}{\bibfnamefont{G.}~\bibnamefont{{Guerrero}}},
  \bibnamefont{and}
  \bibinfo{author}{\bibfnamefont{A.}~\bibnamefont{{Brandenburg}}},
  \bibinfo{journal}{MNRAS} \textbf{\bibinfo{volume}{429}},
  \bibinfo{pages}{1686} (\bibinfo{year}{2013}).

\bibitem[{\citenamefont{{Shapovalov} and {Vishniac}}(2011)}]{shapovalov2013}
\bibinfo{author}{\bibfnamefont{D.~S.} \bibnamefont{{Shapovalov}}}
  \bibnamefont{and} \bibinfo{author}{\bibfnamefont{E.~T.}
  \bibnamefont{{Vishniac}}}, \bibinfo{journal}{ApJ}
  \textbf{\bibinfo{volume}{738}}, \bibinfo{eid}{66} (\bibinfo{year}{2011}).

\bibitem[{\citenamefont{{Lesur} and {Ogilvie}}(2008)}]{lesurdynamo}
\bibinfo{author}{\bibfnamefont{G.}~\bibnamefont{{Lesur}}} \bibnamefont{and}
  \bibinfo{author}{\bibfnamefont{G.~I.} \bibnamefont{{Ogilvie}}},
  \bibinfo{journal}{Astronomy \& Astrophysics} \textbf{\bibinfo{volume}{488}},
  \bibinfo{pages}{451} (\bibinfo{year}{2008}), \eprint{0807.1703}.

\bibitem[{\citenamefont{{Davis} et~al.}(2010)\citenamefont{{Davis}, {Stone},
  and {Pessah}}}]{stratified2010}
\bibinfo{author}{\bibfnamefont{S.~W.} \bibnamefont{{Davis}}},
  \bibinfo{author}{\bibfnamefont{J.~M.} \bibnamefont{{Stone}}},
  \bibnamefont{and} \bibinfo{author}{\bibfnamefont{M.~E.}
  \bibnamefont{{Pessah}}}, \bibinfo{journal}{ApJ}
  \textbf{\bibinfo{volume}{713}}, \bibinfo{pages}{52} (\bibinfo{year}{2010}).

\bibitem[{\citenamefont{{Simon} et~al.}(2011)\citenamefont{{Simon}, {Hawley},
  and {Beckwith}}}]{simon2011}
\bibinfo{author}{\bibfnamefont{J.~B.} \bibnamefont{{Simon}}},
  \bibinfo{author}{\bibfnamefont{J.~F.} \bibnamefont{{Hawley}}},
  \bibnamefont{and}
  \bibinfo{author}{\bibfnamefont{K.}~\bibnamefont{{Beckwith}}},
  \bibinfo{journal}{ApJ} \textbf{\bibinfo{volume}{730}}, \bibinfo{eid}{94}
  (\bibinfo{year}{2011}).

\end{thebibliography}
\end{document}